\documentclass[AMA,STIX1COL]{WileyNJD-v2}

\articletype{Research Article}%

\received{9 April 2023}
\revised{XXX 2023}
\accepted{XXX 2023}

\usepackage{graphicx,psfrag,epsf}
\usepackage{enumerate}
\usepackage{url} 
\usepackage{booktabs,threeparttable}
\usepackage{bbm}
\usepackage{mathtools}

\def\bfA{{\ensuremath{\bf A}}}

\def\bfC{{\ensuremath{\bf C}}}
\def\bfD{{\ensuremath{\bf D}}}

\def\bfH{{\ensuremath{\bf H}}}
\def\bfI{{\ensuremath{\bf I}}}

\def\bfM{{\ensuremath{\bf M}}}

\def\bfR{{\ensuremath{\bf R}}}
\def\bfS{{\ensuremath{\bf S}}}

\def\bfV{{\ensuremath{\bf V}}}
\def\bfW{{\ensuremath{\bf W}}}
\def\bfX{{\ensuremath{\bf X}}}
\def\bfY{{\ensuremath{\bf Y}}}
\def\bfZ{{\ensuremath{\bf Z}}}
\def\bfzero{{\ensuremath{\bf 0}}}

\def\bftheta{{\ensuremath\boldsymbol{\theta}}}

\def\bfalpha{{\ensuremath\boldsymbol{\alpha}}}
\def\bfbeta{{\ensuremath\boldsymbol{\beta}}}
\def\bfgamma{{\ensuremath\boldsymbol{\gamma}}}

\def\bfrho{{\ensuremath{{\boldsymbol{\rho}}}}}

\def\bfLambda{{\ensuremath\boldsymbol{\Lambda}}}
\def\bfmu{{\ensuremath{{\boldsymbol{\mu}}}}}

\def\bfepsilon{{\ensuremath{{\boldsymbol{\epsilon}}}}}

\def\bfpi{{\ensuremath\boldsymbol{\pi}}}

\def\red{\textcolor{black}}

\def\revisionone{\textcolor{black}}

\begin{document}

\title{Analyzing Risk Factors for Post-Acute Recovery in Older Adults with Alzheimer's Disease and Related Dementia: A New Semi-Parametric Model for Large-Scale Medicare Claims}

\author[1]{Biyi Shen}
\author[2]{Haoyu Ren}
\author[3]{Michelle Shardell}
\author[4]{Jason Falvey*}
\author[3]{Chixiang Chen*}
\authormark{B. Shen \textsc{et al}}

\address[1]{\orgname{Regeneron Pharmaceuticals}, \orgaddress{\state{New Jersey}, \country{U.S.A.}}}

\address[2]{\orgdiv{Department of Mathematics and Statistics}, \orgname{University of Maryland, Baltimore County}, \orgaddress{\state{Maryland}, \country{U.S.A.}}}

\address[3]{\orgdiv{Department of Epidemiology and Public Health}, \orgname{University of Maryland School of Medicine}, \orgaddress{\state{Maryland}, \country{U.S.A.}}}

\address[4]{\orgdiv{Department of Physical Therapy and Rehabilitation Science}, \orgname{University of Maryland School of Medicine}, \orgaddress{\state{Maryland}, \country{U.S.A.}}}

\corres{*Chixiang Chen and Jason Falvey, \\ \email{chixiang.chen@som.umaryland.edu}  \\
\email{JFalvey@som.umaryland.edu}}

\abstract[Summary]{Nearly 300,000 older adults experience a hip fracture every year, the majority of which occur following a fall. Unfortunately, recovery after fall-related trauma such as hip fracture is poor, where older adults diagnosed with Alzheimer's Disease and Related Dementia (ADRD) spend a particularly long time in hospitals or rehabilitation facilities during the post-operative recuperation period. Because older adults value functional recovery and spending time at home versus facilities as key outcomes after hospitalization, identifying factors that influence days spent at home after hospitalization is imperative. While several individual-level factors have been identified, the characteristics of the treating hospital have recently been identified as contributors. However, few methodological rigorous approaches are available to help overcome potential sources of bias such as hospital-level unmeasured confounders, informative hospital size, and loss to follow-up due to death. This article develops a useful tool equipped with unsupervised learning to simultaneously handle statistical complexities that are often encountered in health services research, especially when using large administrative claims databases. The proposed estimator has a closed form, thus only requiring light computation load in a large-scale study. We further develop its asymptotic properties with stabilized inference assisted by unsupervised clustering. Extensive simulation studies demonstrate superiority of the proposed estimator compared to existing estimators.}



\maketitle


\section{Introduction}\label{Introduction}

Hip fracture is a common traumatic injury among older adults, with healthcare costs for treatment and recovery exceeding $\$68$ billion per year \cite{center}. Older adults diagnosed with Alzheimer's Disease and related dementia (ADRD) are up to three times more likely than cognitively intact older adults to sustain a hip fracture \cite{friedman2010dementia}. Moreover, patients with ADRD have longer and more expensive recovery trajectories compared to those without ADRD, resulting in a disproportionate economic burden on the healthcare system \cite{friedman2010dementia}.

Given the significant costs and consequences of hip fractures among older adults, improving their post-acute recovery, especially for those with Alzheimer's Disease and related dementias (ADRD), is a crucial national priority. To achieve this objective, a meaningful and objective metric is required that can include additional post-discharge outcomes and assess older adult trauma to evaluate hospital quality and establish targets for best allocation of limited resources to improve future care. A relatively new patient-centered outcome, \red{called "Days At Home" (DAH) calculated as $30$ days minus the number of days spent in hospitals, nursing homes, or rehabilitation facilities in each month}, may serve as a useful marker of patient recovery, which can be entirely derived from Medicare administrative claims \cite{groff2016days,burke2020healthy}. Recent literature has shown that DAH can measure distinct domains of trauma hospital quality, which are different from the mortality rates used historically as a quality benchmark \cite{zogg2022beyond}.
In this article, we aim to further investigate the Medicare Claims database to identify \revisionone{clinically meaningful patient-specific characteristics} highly associated with monthly DAH after hospital discharge during the first half-year of follow-up, to inform personalized care, particularly for older adults with ADRD, whose recovery following hip fractures is generally poorer and not well understood. 

There are multiple statistical challenges when analyzing DAH in Medicare administrative healthcare claims. In this article, we address three potential sources of bias. First, older adults hospitalized with a hip fracture have a high risk of mortality after hospital discharge. As a result of truncation by death, some older adults' observed DAH trajectories will be shorter than others. This truncation by death could be informative due to high association of mortality and DAH \cite{zogg2022beyond}. We note that besides death, any other potential sources of older adult dropout are likely to be negligible in Medicare claims. Second, the number of older adults in each hospital is not fixed and could be informative. Each hospital is a cluster where the outcome of a cluster member given exposure and other covariates might be associated with the exposure levels of other cluster members. That is, the cluster size may confound the relationship between exposure and outcome, a phenomenon called Informative Cluster Size (ICS) \cite{seaman2014review}. Lastly, characteristics of the treating hospital, including geographical region, could be contributors to both patient-level risk factors and DAH. However, such information is not sufficiently available in Medicare Claims. That is, the analysis might be affected by unmeasured hospital-level confounding, which also causes ICS. Any of these three issues will lead to invalid statistical inference \cite{robins1995analysis,pavlou2021risk,shen2022semiparametric}. \revisionone{Our goal is to develop a statistical method that can simultaneously address all three challenges}.

In the presence of truncation by death, consistent estimation can be achieved using the (generalized) linear mixed-effect (LME) model or weighted generalized estimating equation (WGEE) under the assumption of missing at random (MAR) \cite{liang1986longitudinal,robins1995analysis,casals2014methodological,shardell2018joint, chen2019empirical,chen2021multiple}. MAR assumes that truncation is not random, but can be fully accounted for by completely observed variables \cite{bhaskaran2014difference}. In contrast, missing not at random (MNAR) occurs when the truncation relies on variables that are not observable \cite{resseguier2011sensitivity}. MNAR typically requires stronger assumptions for parameter identification that may not be testable in practice \cite{shen2021joint}.  We will assume MAR in our application since DAH observed before death has been demonstrated as a strong indicator of post-discharge mortality \cite{zogg2022beyond}. In the presence of both truncation by death and ICS, within-cluster resampling (WCR) and clustered weighted generalized estimating equation (CWGEE) are recommended, as they have been shown to be asymptotically equivalent \cite{hoffman2001within,williamson2003marginal,wang2011inference,shardell2015doubly,shen2022semiparametric}. Joint modeling (JM) of patient outcome and cluster size is another method that answers the question of how cluster size is associated with the outcome, but it is fully parametric \cite{kong2018conditional,shen2021joint}. In the presence of hospital-level unmeasured confounding, (generalized) LME can mitigate bias by including latent variables representing multi-layer variabilities \cite{casals2014methodological}. Other methods are available to alleviate unmeasured confounding, such as methods based on instrumental variables and negative control variables \cite{angrist2001instrumental,lipsitch2010negative}.

 Despite the existing literature, none of the previously proposed methods can simultaneously address all three methodological challenges that commonly co-occur in health services research involving large claims databases: truncation by death, informative cluster size, and unmeasured confounding at the hospital level. In this article, we present a \revisionone{new} tool to handle these statistical complexities. \revisionone{We first translate the problem of unmeasured hospital confounding to the setup of nuisance parameters, and then apply profile least square to deal with many nuisance parameters and creatively adapt inverse probability weighting to handle dropout missingness and ICS, a first use of combining these methods in literature.}  The proposed estimator has a closed form, making it computationally efficient even in large-scale studies. \revisionone{We derive its asymptotic properties accounting for estimation uncertainty of the inverse probability weighting, which can be used for statistical inference in practice}. To further improve the stability of the estimation of standard error and coverage probability, we employ an unsupervised clustering algorithm for making more accurate statistical inference, \revisionone{which is evaluated through extensive simulation studies}. We then apply the proposed estimator to identify the risk factors associated with DAH among older adults \revisionone{living} with ADRD who have experienced hip fracture, stratified by observed hospital characteristics. \revisionone{This application will help detect high-risk sub-populations and possibly inform hospital resource allocation, thus promoting the advancement of precision healthcare}. Finally, we extend the proposed estimator to handle \revisionone{time-varying unmeasured hospital-level confounders/effects} and high-dimensional \revisionone{patient-level} covariates and discuss future research directions.

The remaining article is organized as follows. Section \ref{method} describes existing methods and the proposed method. Section \ref{Simulation} provides extensive numerical evaluations of the proposed estimator. Section \ref{A real data application} illustrates the application of our method in health service research. Section \ref{discussion} provides the method extension and summarizes the manuscript. All technical proofs, extra simulations, and extensions/more discussions can be found in the Supplementary Material.

\section{Method}\label{method}
\subsection{Notation}\label{notation}
We consider the case of three-level data: hospital-level, patient-level within hospital, and observation-level within patient. Let $i$, $j$, and $k$ be the index of the $i$th hospital, $j$th patient, and $k$th month, respectively. Suppose that there are $N$ hospitals in total in our study; each hospital has specific number of patients; each patient has at most $K$ observations in total after hip fracture. Specifically, there are $n_i$ subjects in the $i$th hospital, for $i=1,\ldots,N$.  Let $\bfY_{ij}=(Y_{ij1}, \ldots, Y_{ijK})^T$ be the vector containing patient-level outcomes, where $Y_{ijk}$ is the $k$th observation of the $j$th patient in the $i$th hospital. Let $\bfR_{ij}=(R_{ij1},\ldots,R_{ijK})^T$ be a vector containing observation indicator, in which $R_{ijk}$ equals $1$ if the $j$th patient has the $k$th observation and equals $0$ if the patient is dead at the $k$th observation. In addition, let the total patient-level sample size be $N_{sub}=\sum_{i=1}^N n_i$; let $\bfX_{ij}=(\bfX^T_{ij1},\ldots,\bfX^T_{ijK})^T$ be a $K\times p$ matrix containing observed patient-level covariates from the $j$th patient at the $i$th hospital (potentially including interactions between patient-level variables and observed hospital-level variables); Similarly, let $\bfM_{i}$ be a $K\times q$ matrix containing unobserved hospital-level covariates from the $i$th hospital.  \revisionone{As our primary focus revolves around post-fracture outcomes following hospital discharge in the data application, we also make the assumption that the variables within $\mathbf{M}_{i}$ should remain the same across all patient observations. We provide an extension for broader applications to address time-varying unmeasured hospital-level confounders in the Supplementary Material.} Thus, the \revisionone{true} model of outcomes $\bfY_{ij}$ is defined as follows
\begin{equation}\label{true model}
   \bfY_{ij} = \bfX_{ij}\bfbeta_0 + \bfM_i\bfgamma_0 + \bfepsilon_{ij}
\end{equation}
 \revisionone{with the true conditional mean $E(\bfY_{ij}|\bfX_{ij},\bfM_{i})=\bfX_{ij}\bfbeta_0+\bfM_{i}\bfgamma_0$}, where $\bfepsilon_{ij}=(\epsilon_{ij1},\ldots,\epsilon_{ijK})^T$ is an independent and identically distributed residual vector. We do not impose any parametric distribution assumption on the residual; $\bfbeta_0$ and $\bfgamma_0$ are \revisionone{the true} parameter vectors, the former of which has length $p$ and is of our primary interest, while the latter of which has length $q$ and quantifies confounding effects. In this true model, the parameters of interest in $\bfbeta_0$ quantify association effects of patient-level factors on DAH for immortal cohort \cite{kurland2009longitudinal}. To make these parameters identifiable, we implicitly assume that there is no interaction between observed and unobserved variables, which is a reasonable assumption for our data application. We consider the linear model in our application as it has a more straightforward interpretation and stable computation. However, we emphasize that our model does not impose any parametric likelihood assumption, such as normally distributed residuals. Throughout the remaining article, we assume a fixed number of parameters and that the covariates are predetermined by domain experts. The extension to the case with high-dimensional covariates is presented in Section \revisionone{3} of the Supplementary Material. For the purpose of model development, we further use $\bfY=(\bfY_{11},\ldots, \bfY_{N1},\ldots)^T$, $\bfX=(\bfX_{11}^T,\ldots,\bfX_{N1}^T,\ldots)^T$, $\bfepsilon=(\bfepsilon^T_{11},\ldots,\bfepsilon^T_{N1},\ldots)^T$, and $\bfR=(\bfR^T_{11},\ldots,\bfR^T_{N1},\ldots)^T$  to denote the stacked outcomes, observed covariates,  residuals, and observation indicators, respectively.

\subsection{Existing method}\label{Existing method}

\subsubsection*{Generalized estimating equation (GEE)\\}

GEE is a robust semi-parametric approach that produces marginal inference on parameter estimates \cite{zeger1986longitudinal} and allows for misspecification of the correlation structure of responses given mild regularity conditions. 
The link function $l(\cdot)$ is known and pre-specified depending on the outcome type, satisfying $l(\bfmu_{ij})=\bfX_{ij} \bfbeta$ with $\bfmu_{ij}=E(\bfY_{ij}|\bfX_{ij})$.
Note that the variance-covariance matrix $\bfV_{ij}$ can be written as $\bfV_{ij}=\bfA_{ij}^{1/2}\bfC_{ij} (\bfrho) \bfA_{ij}^{1/2}$, where $\bfA_{ij}$ is a diagonal matrix containing variance components, and $\bfC_{ij} (\bfrho)$ is a prespecified ``working" correlation matrix depending on a set of parameters $\bfrho$, which describes the association pattern of within-cluster measurements and should be estimated. Applying GEE to clustered data, the estimate $\hat{\bfbeta}_{GEE}$ is obtained by solving the following estimating equations
\begin{equation*}
 \sum_{i=1}^N\sum_{j=1}^{n_i}\bfD_{ij}^T \bfV_{ij}^{-1} (\bfY_{ij}-\bfmu_{ij})=\bfzero,
\end{equation*}
with $\bfD_{ij}=\partial \bfmu_{ij} /{\partial \bfbeta^T}$. We remark here that GEE is valid only if the data are complete or follow missing completely at random mechanism and cannot handle the issues of ICS and unmeasured confounding in the main model.

\subsubsection*{Weighted GEE (WGEE)\\}

WGEE was initially proposed to handle {missing data} under the MAR assumption by utilizing inverse probability weighting (IPW) \cite{robins1995analysis}. Specifically, let $\bfpi_{ij}=(\pi_{ij1}, \ldots, \pi_{ijK})^T$ be the probability vector of observing the outcome for the $j$th subject in the $i$th hospital, where $\pi_{ijk}=Pr(R_{ijk}=1| \tilde\bfH_{ijk})$, and $\tilde\bfH_{ijk}$ is the vector consisting of all observed variables until time $k$ including observed outcomes. Moreover, $\bfpi_{ijk}=\lambda_{ij1} \times \lambda_{ij2}\times \dots \times \lambda_{ijk}$ where $\lambda_{ij1}=1$ (the outcome at baseline is always observed) and $\lambda_{ijk}=Pr(R_{ijk}=1|R_{ij(k-1)}=1, \tilde\bfH_{ijk};\bfalpha)$, for $k=2, \ldots, K$, are conditional models given previous observation status parameterized by $\bfalpha$, which can be modeled by logistic regression. Note that we impose $\lambda_{ijk}=Pr(R_{ijk}=1|R_{ij(k-1)}=0, \tilde\bfH_{ijk};\bfalpha)=0$ owing to death as an absorbing state. Finally, by defining a diagonal weight matrix ${\bfW}_{ij}({\bfalpha})=Diag(\bfR_{ij}/\bfpi_{ij})$, the parameter estimate, denoted as $\hat{\bfbeta}_{WGEE}$, can be obtained by solving the estimating equations
\begin{equation*}
\sum_{i=1}^{N}\sum_{j=1}^{n_i}\bfD_{ij}^T \bfV_{ij}^{-1}\hat{\bfW}_{ij} (\bfY_{ij}-\bfmu_{ij})=\bfzero,
\end{equation*}
where $\hat{\bfW}_{ij}$ is equal to ${\bfW}_{ij}$ by replacing $\bfalpha$ with an estimate $\hat{\bfalpha}$ by maximizing partial likelihood \cite{chen2019empirical}. The estimation of $\bfbeta$ is unbiased if the probability of observing the outcome is correctly modeled. Like GEE, however, WGEE cannot handle the issues of ICS and unmeasured confounding in the main model.

\subsubsection*{Cluster-weighted GEE (CWGEE)\\}

CWGEE\citep{williamson2003marginal} is another marginal approach to handle the issue of ICS in a clustered analysis. The extension to longitudinal clustered data (possibly with informative dropout over time) was investigated by more recent literature \citep{wang2011inference,mitani2022accounting}. This method uses the number of patients in hospital as extra weight in WGEE. 
Thereby, CWGEE utilizes the following estimating equations
\begin{equation*}
\sum_{i=1}^{N}\frac{1}{n_i}\sum_{j=1}^{n_i}\bfD_{ij}^T \bfV_{ij}^{-1}\hat{\bfW}_{ij} (\bfY_{ij}-\bfmu_{ij})=\bfzero.
\end{equation*}
Despite its advantage over WGEE, CWGEE cannot handle the issue of unmeasured confounding in the main model.

\subsubsection*{Mixed-effect model (MM)\\}

MM is a popular choice for modeling a hierarchical and complex data structure. In our application, month-level observations are nested in subjects that are nested in hospitals. The three-level mixed-effects model with random intercepts at both subject-level and hospital-level can be written as
\begin{equation}
    Y_{ijk}=\bfX^T_{ijk}{\bfbeta}+\gamma_i+\nu_{ij}+\epsilon_{ijk}.
\end{equation}
The hospital-level random effect $\gamma_i$ indicates the influence of hospital $i$ on the measured outcome,
while the subject-level random effect $\nu_{ij}$ represents the influence of subject $j$ from hospital $i$. The
population distributions of $\gamma_i$ and $\nu_{ij}$ are often assumed to be normal. More complex models can be considered by incorporating additional random effects that interact with observed covariates or even are linked to the death (survival) model \cite{shen2021joint}. While the MM-based approach has the potential to address the issues of truncation by death under the MNAR assumption, ICS, and \revisionone{hierarchical data structure}, it relies on the assumption that the underlying truth precisely follows the specified parametric model. This may not be unlikely in large observational studies \revisionone{and with unmeasured hospital-level confounding}, and the computational complexity of the MM approach can lead to difficulties in convergence when multiple random effects are involved. These limitations are demonstrated in our simulation studies.

\subsection{Proposed method}
This section outlines our proposed semi-parametric estimator, which is computationally convenient and relies less on parametric assumptions. Let us revisit the true model in (\ref{true model}) and first consider the issue of hospital-level confounding. Since the variables in $\bfM_i$ cannot be observed, regressing outcomes $\bfY_{ij}$ only on the observed variables $\bfX_{ij}$ will lead to biased estimates of $\bfbeta$. To overcome this issue, we first reformulate the model in (\ref{true model}) by introducing an observed matrix and nuisance parameter vector and using the stacked notations in Section \ref{notation}, i.e.,
\begin{equation}\label{eq1}
   \bfY = \bfX\bfbeta + \bfZ\bftheta + \bfepsilon,
\end{equation}
    where $\bfZ=(\bfZ_1,\ldots,\bfZ_{N})$ is a $KN_{{sub}}\times N$ matrix consisting of hospital-specific intercepts, i.e., the vector $\bfZ_{i}$ is the indicator of the $i$th hospital, of which the element equals $1$ if the corresponding observation is from the $i$th hospital and equals $0$, otherwise. The $\bftheta$ is a $N\times 1$ vector containing nuisance parameters. \revisionone{To establish a linkage to the true conditional mean in (\ref{true model}) and achieve a consistent estimate of $\bfbeta_0$, we make the assumption that, conditional on $\bfM$ and $\bfZ$, the $i$th element of the underlying truth $\bftheta_0$ in the model (\ref{eq1}) is equal to the unique element in $\bfM_{i}\bfgamma_0$ in the true conditional mean.}
\revisionone{Intuitively, the reformulated model in (\ref{eq1}) treats unobserved components in (\ref{true model}) as saturated nuisance parameters, and thus, by estimating these nuisance parameters could help recover the hospital-level unmeasured component and resolve the issue of biased estimation for $\bfbeta$.} However, \revisionone{without the knowledge of hospital-level latent confounders in $\bfM_{i}$, all elements in $\bftheta$ should be estimated, and} the dimension of $\bftheta$ can be very high in practice where there are over a thousand hospitals in Medicare claims. Estimating high-dimensional nuisance parameters may lead to \revisionone{intractable or less desirable statistical
inference for the main estimates $\bfbeta$ and may yield unstable numerical performance} \cite{portnoy1984asymptotic,portnoy1985asymptotic, fan2004new,ning2017general,battey2020high}. 

To develop a valid and computationally scalable estimator, we adopt the idea of the profile least square (PLS), which is widely used in the context of partial linear models and non-parametric regression \cite{fan2004new,balabdaoui2020profile}. The procedure is summarized below: we first rewrite the equation in (\ref{eq1}) by $\bfY - \bfX\bfbeta = \bfZ\bftheta + \bfepsilon$ and pretend that $\bfbeta$ parameter values are known. Then, the parameter vector $\bftheta$ can be estimated by a standard regression. We denote the resulting ``estimate" as $\hat{\bftheta}(\bfbeta)$, which is a function of the primary parameter vector. However, such an ``estimate" is not calculable, owing to unknown vector $\bfbeta$. Thereby, we plug back the ``estimate" $\hat{\bftheta}(\bfbeta)$ to the model in (\ref{eq1}) and then estimate $\bfbeta$. Moreover, in the presence of ICS and patient truncation by death, the above procedure can be adapted as follows: first, obtain the $\bftheta$ estimate as if the $\bfbeta$ is known by solving a weighted estimating equations
$\bfZ^{T}\hat{\bfW}(\bfY - \bfX\bfbeta - \bfZ\bftheta) = \bfzero$, where $\hat{\bfW}$ is a $KN_{sub}\times KN_{sub}$ weight matrix used to address bias owing to truncation by death and ICS. We consider two candidate forms:  $\hat{\bfW}_1=Diag({\bfR}/{\hat{\bfpi}})$  and $\hat{\bfW}_2=Diag({\bfR}/{\hat{\bfpi}^\ast})$,  where the estimated probability vector $\hat{\bfpi}=(\hat{\bfpi}_{11}^T,\ldots,\hat{\bfpi}_{N1}^T,\ldots)^T$ is defined in Section \ref{Existing method} and solved by maximizing the partial likelihood \cite{chen2019empirical}, and $\hat{\bfpi}^\ast=(n_1\hat{\bfpi}_{11}^T,\ldots,n_N\hat{\bfpi}_{N1}^T,\ldots)^T$. Similar to WGEE, the term ${\bfR}/{\hat{\bfpi}}$ in $\hat{\bfW}_1$ is used to address the issue of truncation by death (here, we assume the MAR assumption and that the probability of observing the outcome is correctly modeled. The violation of this assumption is discussed in Section \ref{discussion}); and similar to CWGEE, the extra weight $n_i$ in $\hat{\bfW}_2$ is to adjust the issue of informative hospital size. We remark here that, under the common assumption (4) in Section \ref{Asymptotic property}, the issue of ICS is indeed attributed to the issue of unmeasured hospital-level confounding. Thus, addressing the latter issue will automatically resolve the former \revisionone{one}, which implies the extra weight $n_i$ may not be necessary, and thereby the weight $\hat{\bfW}_1$ is sufficient. Based on numerical evaluations in Section \ref{Simulation}, we find that two types of weights lead to little estimation bias, whereas the weight $\hat{\bfW}_1$ leads to a bit better estimation of asymptotic standard error. By using either $\hat{\bfW}_1$ or $\hat{\bfW}_2$ and solving the estimating equations above, we have
\begin{equation}\label{theta estimate}
    \hat{\bftheta} = \hat{\bfS}(\bfY - \bfX\bfbeta),
\end{equation}
where $\hat{\bfS} = [\bfZ^{T}\hat{\bfW}\bfZ]^{-1}\bfZ^{T}\hat{\bfW}$, \revisionone{a block-wise matrix and easy to calculate}. Plugging this estimator back into model (\ref{eq1}), we have  
\begin{equation}\label{beta estimate}
    \hat{\bfbeta} = \{\bfX^{T}(\bfI - \bfZ\hat{\bfS})^{T}\hat{\bfW}(\bfI - \bfZ\hat{\bfS})\bfX\}^{-1}\bfX^{T}(\bfI - \bfZ\hat{\bfS})^T\hat{\bfW}(\bfI - \bfZ\hat{\bfS})\bfY.
\end{equation}

\revisionone{The proposed estimator differs significantly in both theory and application from the traditional PLS estimator \cite{fan2004new}.} The traditional estimator considers a partial linear model with a non-parametric component and a fixed weight matrix $\bfW$, while our estimation focuses on addressing unmeasured confounding and relying on the estimated weight matrix using IPW methods, as shown in (\ref{beta estimate}). \revisionone{This leads to a different asymptotic variance by accounting for extra variability from IPW estimation uncertainty and thus affects the statistical inference}. Other unique features of our estimator (\ref{beta estimate}) are worth noting. First, the estimator (\ref{beta estimate}) has a closed form, which is computationally efficient and numerically stable for analyzing large-scale Medicare Claims; second, nuisance parameters $\bftheta$ are profiled out by PLS, thus the resulting estimator is not sensitive to \revisionone{hospital-specific unmeasured effects}; third, estimation simultaneously addresses unmeasured hospital-level confounding, ICS, and truncation by death. In Section \ref{Asymptotic property}, we demonstrate that our proposed estimator is consistent and follows an asymptotic normal distribution. To further stabilize the inference, we augment the estimation with unsupervised clustering, as described in more detail in Section \ref{Asymptotic property}. Due to its flexibility in handling statistical challenges in analyzing large-scale Medicare claims, we refer to the proposed estimator as an omnibus estimator (Omni).

We notice here that the independence ``working" correlation structure is implicitly assumed in our estimation. Similar to GEE and WGEE, the specification of ``working" correlation structure will not affect the consistency of the proposed estimator, but it may affect the estimation efficiency \cite{liang1986longitudinal, robins1995analysis}. However, two facts support the use of independence structure: first, similar to WGEE with inverse probability weighting involved in the estimation, correctly specifying the correlation structure may not guarantee higher efficiency \cite{han2015achieving}; second, \revisionone{efficiency itself is not a primary concern in our data application since we have more than ten thousands of patients. Thus, developing a computationally efficient estimator with small estimation bias is the main target in this article}. Moreover, we have extended our estimator to the application of \revisionone{high dimensional patient-level variables and time-varying hospital-level unmeasured effects}. Please refer to Section \ref{discussion} and the Supplementary Material for more details. 

\subsection{Asymptotic properties}\label{Asymptotic property}
This section illustrates asymptotic properties of the estimator in (\ref{beta estimate}) and proposes one procedure for stabilizing statistical inference. To facilitate theoretical derivations, we require the following three assumptions:
(i) no interaction between observed patient-level variables and unmeasured hospital-level variables; (ii) the model of truncation by death is correctly specified, and the probabilities $\pi_{ijk} >0$ for all $i,j,k$; (iii) the estimator $\hat{\bftheta}$ in (\ref{beta estimate}) consistently estimates the true $\bftheta$; (iv) \revisionone{we assume conditional independence
between the outcome and the number of patients in hospital given all patient-level and hospital-level covariates}. The first assumption guarantees parameter identifiability in the presence of unmeasured hospital-level confounders. This assumption is reasonable in our application, as there is little evidence in the literature supporting the existence of interaction effects on hip-fracture recovery from hospital-level variables and patient-level variables. The second assumption is necessary for validly using inverse probability weighting, which is common in the literature \cite{robins1994estimation,robins1995analysis,chen2019empirical}. Violation of this assumption is discussed in Section \ref{discussion}. The third assumption typically holds in practice, given a reasonable number of patients in each hospital. Violation of this assumption, such as having only a few patients in each hospital, is investigated at the end of this section. \revisionone{The last assumption is more like common sense in the literature, pointing out the root of ICS, and crucial for the valid application of the inverse cluster size weighting technique \cite{wang2011inference, seaman2014review}}.

With the assistance of well-recognized regular moment conditions \cite{newey1994large}, we have the following theorem
\begin{theorem}\label{thm1}
    The proposed $\hat{\bfbeta}$ in (\ref{beta estimate}) is a consistent estimator of $\bfbeta_0$, and $\sqrt{N_{sub}}(\hat{\bfbeta}-\bfbeta_0)$ has a normal distribution with a mean-zero vector and variance-covariance matrix
   ${\bfD}^{-1}({\bfV}_1 - {\bfLambda}\bfV_2\bfLambda^T)({\bfD}^{-1})^T$,
as $N_{sub}$ goes to infity.
The details of ${\bfD}$, ${\bfV}_1$, ${\bfV}_2$, and ${\bfLambda}$ are described in the Supplementary Material.
\end{theorem}
The variance can be estimated by replacing ${\bfD}$, ${\bfV}_1$, ${\bfV}_2$, and ${\bfLambda}$ with their \revisionone{sample moment} estimators $\hat{\bfD}$, $\hat{\bfV}_1$, $\hat{\bfV}_2$, and $\hat{\bfLambda}$, respectively. The variance structure in Theorem \ref{thm1} is different from the one in literature \citep{fan2004new} in the sense that it involves extra variability from estimating IPW. 

\begin{figure}
    \centering
    \includegraphics[width=4in]{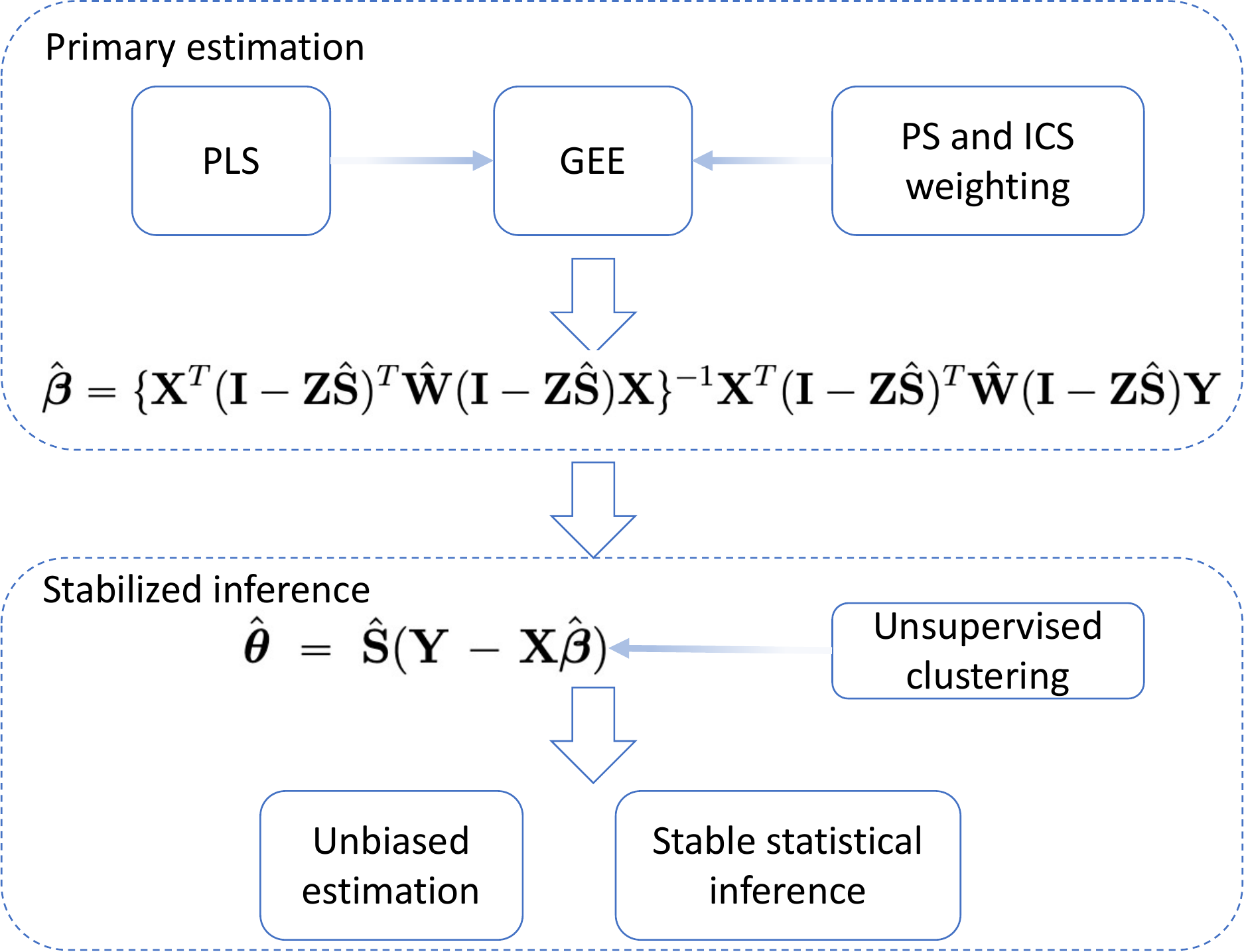}
    \caption{A schematic workflow of the proposed method. The proposed framework is robust and computationally efficient and leads to reliable statistical inference. PLS: profile least square; PS: propensity score; ICS: informative cluster size.}
    \label{method workflow}
\end{figure}

\textbf{Unsupervised clustering-assisted inference}. Though the estimator $\hat{\bfbeta}$ in (\ref{beta estimate}) is not sensitive to the estimated nuisance parameters in $\hat{\bftheta}$ via (\ref{theta estimate}) by replacing ${\bfbeta}$ with $\hat{\bfbeta}$, estimation of $\bfV_1$ in the variance-covariance matrix is sensitive to the nuisance parameter estimation, especially in a situation where there are numerous hospitals but only a few patients within each hospital. Under this case, the regular large sample theory may not hold for $\hat{\bftheta}$, and the resulting estimates can substantially vary across hospitals. An inaccurate estimate of $\hat{\bftheta}$ may lead to less accurate estimate of variance-covariance matrix of $\hat{\bfbeta}$, thus less reliable statistical inference. To improve statistical inference, we further adopt unsupervised clustering algorithm to stabilize $\bftheta$ estimation, such as k-Means, Louvain, DBSCAN, etc \cite{pham2005selection,jang2018dbscan++}.  Specifically, after calculating \revisionone{the estimates} $\hat{\bftheta}=\hat{\bfS}(\bfY-\bfX\hat{\bfbeta})$, we further apply an unsupervised clustering algorithm to capture latent patterns among \revisionone{these} estimates. By doing so, we replace substantially perturbed $\bftheta$ estimates by centroids of resulting clusters, thus stabilizing the estimation of high-dimensional nuisance parameters. The number of latent clusters should be determined. \revisionone{We employ a
procedure to automatically determine the most suitable cluster size: for instance, given a predefined set of cluster sizes, we
conduct K-means clustering for $\hat{\bftheta}$ and calculate the Silhouette score for each cluster size. The ”optimal” cluster size is defined as the one corresponding to the highest Silhouette score\cite{rousseeuw1987silhouettes, pham2005selection}, indicating the best separation.
Subsequently, we use the clustering results based on the selected cluster size to compute asymptotic standard errors.} The sketched numerical procedure is summarized by Figure \ref{method workflow}. The underlying rationale for applying clustering of $\hat{\bftheta}$ in practice is the plausibility that many hospitals may share similar unobserved features \revisionone{and produce comparable effects on post-fracture recovery outcomes. This similarity may arise from shared geographic factors and healthcare policies. As a result, the elements in the vector $\bftheta$ can be grouped into clusters, each having unique effects on outcomes, and only a limited number of unique values will be present in $\bftheta$, analogous to the sparsity assumption in the setting of high-dimensional variable selection \cite{wasserman2009high}.} Thus, machine learning-based unsupervised \revisionone{clustering} can borrow information across hospitals and aggregate similar $\hat{\bftheta}$ into a cluster. We also notice that clustering hospital-level intercepts also help delineate heterogeneity profiles among hospitals, which has its own clinical meaning. We will illustrate an example in the simulation section that shows the significant benefits we could gain from the machine learning-based algorithm when the hospital number is large, but each hospital has a very small sample size.

\begin{table}
\centering
\caption{Evaluation of bias under truncation rate around $30\%$ and four combinations of hospital number and patient number in hospital\red{, with residuals following normal distribution}. The true values of parameters are $\beta_0=1$,  $\beta_1=0.5$, $\beta_2=-0.5$, $\beta_3=0.5$, $\beta_4=1$.  MM: mixed-effect model; Omni: our proposed method. \red{All the values listed here are multiplied by $100$.}}
\label{biashigh}
\begin{tabular}{lcccccc}
\toprule
30\% truncation rate                   &            & MM  & GEE   & WGEE  & CWGEE & Omni \\
\midrule
100 hospitals                  & $x_1$ & 2.95  & 17.50 & 13.09 & 15.54 & 0.01 \\
$\sim$60 subjects per hospital & $x_2$ & 0.75  & 4.73  & 3.73  & 3.89  & 0.04 \\
                               & $x_3$ & 4.47  & 53.29 & 57.44 & 54.77 & 0.24 \\
                               & $x_4$ & 0.67  & 1.43  & 2.71  & 0.82  & 0.16 \\
                               & $x_5$ & 1.64  & 3.11  & 5.93  & 1.62  & 0.09 \\\midrule
100 hospitals                  & $x_1$ & 10.13 & 17.37 & 13.12 & 15.72 & 0.11 \\
$\sim$25 subjects per hospital & $x_2$ & 2.85  & 4.81  & 3.64  & 4.08  & 0.07 \\
                               & $x_3$ & 17.98 & 53.45 & 57.33 & 54.21 & 0.28 \\
                               & $x_4$ & 2.56  & 1.44  & 2.61  & 0.61  & 0.19 \\
                               & $x_5$ & 5.11  & 2.84  & 5.79  & 1.87  & 0.40 \\ \midrule
25 hospitals                   & $x_1$ & 3.12  & 17.40 & 13.51 & 15.59 & 0.08 \\
$\sim$60 subjects per hospital & $x_2$ & 0.86  & 4.90  & 4.10  & 4.20  & 0.10 \\
                               & $x_3$ & 4.70  & 52.66 & 55.90 & 54.03 & 0.29 \\
                               & $x_4$ & 0.92  & 1.65  & 2.54  & 0.62  & 0.02 \\
                               & $x_5$ & 1.49  & 2.98  & 5.81  & 2.17  & 0.34 \\ \midrule
25 hospitals                   & $x_1$ & 10.25 & 17.40 & 14.12 & 16.10 & 0.30 \\
$\sim$25 subjects per hospital & $x_2$ & 2.70  & 4.69  & 3.76  & 4.04  & 0.24 \\
                               & $x_3$ & 18.75 & 52.45 & 54.52 & 52.80 & 0.19 \\
                               & $x_4$ & 2.71  & 1.71  & 2.55  & 0.46  & 0.07 \\
                               & $x_5$ & 5.33  & 3.25  & 5.35  & 1.12  & 0.07\\\bottomrule
\end{tabular}
\end{table}

\begin{table}
\centering
\caption{Evaluation of bias under truncation rate around $30\%$ and four combinations of hospital number and patient number in hospital\red{, with residual following non-normal and skewed distribution}. The true values of parameters are $\beta_0=1$,  $\beta_1=0.5$, $\beta_2=-0.5$, $\beta_3=0.5$, $\beta_4=1$.  MM: mixed-effect model; Omni: our proposed method. \red{All the values listed here are multiplied by $100$.}}
\label{biashighwithgamma}
\red{
\begin{tabular}{lcccccc}
\toprule
30\% truncation rate                   &            & MM  & GEE   & WGEE  & CWGEE & Omni \\\midrule
100 hospitals 	&	$x_1$	&	2.91	&	17.43	&	12.68	&	15.25	&	0.01	\\
$\sim$60 subjects per hospital	&	$x_2$	&	0.83	&	4.71	&	3.54	&	3.81	&	0.04	\\
	&	$x_3$	&	4.70	&	53.58	&	58.19	&	55.48	&	0.05	\\
	&	$x_4$	&	0.62	&	1.39	&	2.88	&	0.90	&	0.21	\\
	&	$x_5$	&	1.48	&	3.10	&	6.19	&	1.22	&	0.11	\\\midrule
100 hospitals 	&	$x_1$	&	14.54	&	19.41	&	15.42	&	17.10	&	0.07	\\
$\sim$25 subjects per hospital	&	$x_2$	&	4.03	&	5.26	&	4.22	&	4.41	&	0.01	\\
	&	$x_3$	&	22.33	&	50.28	&	53.89	&	52.36	&	0.10	\\
	&	$x_4$	&	2.58	&	0.61	&	1.67	&	1.42	&	0.23	\\
	&	$x_5$	&	5.19	&	1.26	&	3.33	&	2.98	&	0.48	\\\midrule
25 hospitals 	&	$x_1$	&	3.10	&	17.40	&	13.51	&	15.59	&	0.08	\\
$\sim$60 subjects per hospital	&	$x_2$	&	0.84	&	4.90	&	4.09	&	4.20	&	0.10	\\
	&	$x_3$	&	4.74	&	52.66	&	55.90	&	54.03	&	0.29	\\
	&	$x_4$	&	0.96	&	1.65	&	2.54	&	0.62	&	0.02	\\
	&	$x_5$	&	1.53	&	2.98	&	5.82	&	2.17	&	0.34	\\\midrule
25 hospitals 	&	$x_1$	&	13.88	&	19.50	&	15.92	&	17.58	&	0.35	\\
$\sim$60 subjects per hospital	&	$x_2$	&	3.95	&	5.42	&	4.56	&	4.32	&	0.11	\\
	&	$x_3$	&	20.97	&	49.10	&	52.31	&	50.87	&	0.76	\\
	&	$x_4$	&	2.27	&	0.47	&	1.05	&	1.70	&	0.50	\\
	&	$x_5$	&	5.33	&	1.99	&	3.95	&	2.22	&	0.18
\\\bottomrule
\end{tabular}}
\end{table}

\begin{table}
\centering
\caption{Evaluation of MCSD, ASE, and CP under truncation rate around $30\%$. The true values of parameters are $\beta_0=1$,  $\beta_1=0.5$, $\beta_2=-0.5$, $\beta_3=0.5$, $\beta_4=1$. The values under existing methods are MCSD.  MCSD: Monte Carlo standard deviation; ASE: Asymptotic standard error; CP: $95\%$coverage probability.\red{We apply our proposed method by utilizing K-means procedure to automatically cluster $\hat{\theta}$}.\red{All the values listed here are multiplied by 100.} }
\label{sdhigh}
\red{
\begin{tabular}{llccccccc}
\toprule
\multicolumn{2}{l}{Omni with K-means}         & \multicolumn{3}{c}{Non-Normal distributed residual} &  & \multicolumn{3}{c}{Normal distributed residual} \\ \cline{3-5} \cline{7-9} 
\multicolumn{2}{l}{30\% truncation rate} & MCSD       & ASE        & CP       &  & MCSD     & ASE     & CP    \\\midrule
100 hospitals                   & $x_1$  & 1.95       & 1.85       & 95       &  & 1.31     & 1.33    & 95    \\
$\sim$60 subjects per hospital  & $x_2$  & 2.11       & 2.07       & 95       &  & 1.70     & 1.62    & 94    \\
                                & $x_3$  & 1.10       & 1.06       & 94       &  & 2.25     & 2.18    & 94    \\
                                & $x_4$  & 2.28       & 2.11       & 94       &  & 1.14     & 1.12    & 94    \\
                                & $x_5$  & 2.19       & 2.03       & 94       &  & 2.13     & 2.12    & 94    \\\midrule
100 hospitals                   & $x_1$  & 3.05       & 3.01       & 95       &  & 2.42     & 2.25    & 94    \\
$\sim$25 subjects per hospital  & $x_2$  & 3.95       & 3.83       & 94       &  & 2.79     & 2.66    & 94    \\
                                & $x_3$  & 1.98       & 1.93       & 94       &  & 3.89     & 3.62    & 94    \\
                                & $x_4$  & 3.80       & 3.82       & 94       &  & 1.83     & 1.84    & 96    \\
                                & $x_5$  & 3.03       & 2.73       & 93       &  & 3.61     & 3.63    & 95    \\\midrule
25 hospitals                    & $x_1$  & 2.42       & 2.25       & 94       &  & 3.03     & 2.73    & 93    \\
$\sim$60 subjects per hospital  & $x_2$  & 2.79       & 2.66       & 94       &  & 3.47     & 3.23    & 95    \\
                                & $x_3$  & 3.89       & 3.62       & 94       &  & 4.49     & 4.40    & 95    \\
                                & $x_4$  & 1.83       & 1.84       & 96       &  & 2.29     & 2.25    & 96    \\
                                & $x_5$  & 3.61       & 3.63       & 95       &  & 4.53     & 4.40    & 95    \\\midrule
25 hospitals                    & $x_1$  & 5.46       & 5.06       & 94       &  & 4.61     & 4.43    & 95    \\
$\sim$25 subjects per hospital  & $x_2$  & 6.46       & 5.93       & 93       &  & 5.50     & 5.19    & 95    \\
                                & $x_3$  & 7.96       & 7.70       & 94       &  & 7.87     & 7.28    & 94    \\
                                & $x_4$  & 4.38       & 3.86       & 92       &  & 3.91     & 3.65    & 93    \\
                                & $x_5$  & 8.52       & 7.71       & 93       &  & 7.49     & 7.27    & 94 \\ \bottomrule  
\end{tabular}}
\end{table}

\begin{table}
\centering
\caption{Evaluation of Bias MCSD, ASE, and CP without and with utilizing K-means algorithm when hospital numbers are large \red{but each hospital has few subjects (averagely <10 pts per hospital). Silhouette score is used when applying K-means to automatically select the optimal cluster size.} The true values of parameters are $\beta_0=1$,  $\beta_1=0.5$, $\beta_2=-0.5$, $\beta_3=0.5$, $\beta_4=1$. MCSD: Monte Carlo standard deviation; ASE: Asymptotic standard error; CP: $95\%$coverage probability.\red{All the values listed here are multiplied by 100.}  }
\label{kmeans}
\red{
\begin{tabular}{llcccccccc}
\toprule
Omni                 &       &       &      &  & \multicolumn{2}{c}{Without K-Means} &  & \multicolumn{2}{c}{With K-means} \\ \cline{6-7} \cline{9-10} 
30\% truncation rate &       & Bias  & MCSD &  & ASE              & CP               &  & ASE             & CP             \\\midrule
250 hospitals        & $x_1$ & 0.17  & 2.03 &  & 1.94             & 95.2             &  & 2.01            & 95.6           \\
Normal distributed residual  & $x_2$ & -0.07 & 2.13 &  & 2.08             & 93.4             &  & 2.16            & 94.6           \\
                     & $x_3$ & -0.11 & 3.41 &  & 3.16             & 93.0             &  & 3.37            & 95.4           \\
                     & $x_4$ & -0.26 & 1.74 &  & 1.6              & 91.0             &  & 1.72            & 93.2           \\
                     & $x_5$ & -0.58 & 3.43 &  & 3.15             & 91.6             &  & 3.36            & 94.4           \\\midrule
500 hospitals        & $x_1$ & 0.18  & 1.48 &  & 1.37             & 94.4             &  & 1.42            & 95.6           \\
Normal distributed residual  & $x_2$ & 0.12  & 1.50 &  & 1.47             & 94.2             &  & 1.53            & 95.2           \\
                     & $x_3$ & 0.23  & 2.43 &  & 2.24             & 92.4             &  & 2.40            & 94.0           \\
                     & $x_4$ & -0.23 & 1.30 &  & 1.13             & 92.0             &  & 1.23            & 94.0           \\
                     & $x_5$ & -0.52 & 2.39 &  & 2.24             & 91.6             &  & 2.40            & 93.6           \\\midrule
250 hospitals        & $x_1$ & -0.06 & 2.22 &  & 2.19             & 95.2             &  & 2.27            & 96.2           \\
Non-Normal distributed residual   & $x_2$ & 0.23  & 2.33 &  & 2.35             & 95.4             &  & 2.43            & 95.4           \\
                     & $x_3$ & 0.38  & 3.63 &  & 3.32             & 92.0             &  & 3.56            & 94.2           \\
                     & $x_4$ & -0.30 & 1.80 &  & 1.67             & 92.8             &  & 1.79            & 95.0           \\
                     & $x_5$ & -0.64 & 3.59 &  & 3.31             & 92.8             &  & 3.54            & 94.4           \\\midrule
500 hospitals        & $x_1$ & 0.08  & 1.61 &  & 1.56             & 95.4             &  & 1.60             & 95.6           \\
Non-Normal distributed residual   & $x_2$ & 0.04  & 1.78 &  & 1.67             & 94.3             &  & 1.76             & 95.1           \\
                     & $x_3$ & 0.09  & 2.72 &  & 2.36             & 91.6             &  & 2.59             & 93.5           \\
                     & $x_4$ & -0.34 & 1.33 &  & 1.19             & 90.1             &  & 1.30             & 93.8           \\
                     & $x_5$ & -0.58 & 2.67 &  & 2.35             & 91.8             &  & 2.54             & 93.0          \\\bottomrule            
\end{tabular}
}
\end{table}

\section{Simulation}\label{Simulation}
\subsection{Data generation}\label{Data generation}
Given our data application, we generated longitudinal data with continuous measurements for empirical evaluation (the summary of simulation scheme was provided in Figure S1 of the Supplementary Material). In our simulation studies, we aimed to investigate how the cluster size (i.e., the number of subjects in one hospital) and the number of hospitals affected the accuracy and precision of the $\bfbeta$ estimation. Additionally, we evaluated the effectiveness of our proposed method in addressing three statistical challenges: truncation by death, ICS, and unmeasured hospital-level confounders.
 
The overall data generating scheme is summarized below.
For $i$th hospital, suppose there were two underlying unknown hospital-level confounders $M_{1,i}$ and $M_{2,i}$ following $\text{Binomial}( 2, 0.5)$ and $\text{Binomial}(1, 0.5)$, respectively (\text{Binomial}$(a,b)$ stands for a Binomial distribution with size $a$ and success probability $b$), and two known hospital-level covariates $H_{1,i}$  and $H_{2,i} $ following $\text{Normal}(3, 1)$ and $\text{Binomial}(1, 0.5)$, respectively. We noticed here that, for the $j$th subject and the $k$th month in the $i$th hospital, the hospital-level variables were unvarying, i.e., $M_{*,ijk}=M_{*,i}$ and $H_{*,ijk}=H_{*,i}$.  We then generated the cluster size $n_i$ in the $i$th hospital by a Poisson distribution with the mean structure of $\exp(\gamma_0+\gamma_1H_{1,i}+\gamma_2H_{2,i}+\gamma_3M_{1,i})$. We varied the values of $\gamma$ parameters to achieve desired cluster sizes, with $\gamma_0=1.5$, $\gamma_1=0.5$, $\gamma_2=-0.5$, $\gamma_3=-1$ leading to averagely 25 subjects per hospital, and $\gamma_0=0.85$, $\gamma_1=0.5$, $\gamma_2=0.5$, $\gamma_3=1$ leading to averagely 60 subjects per hospital. After generating the informative cluster size per hospital, we generated subject-level data. There were in total six observations for each subject ($K=6$), but death might occur after the first month. Specifically, we first generated the complete data without missingness and assumed there were five subject-level covariates $X_{1,ijk}$, $X_{2,ijk}$, $X_{3,ijk}$, $X_{4,ijk}$, $X_{5,ijk}$, which could be either time-varying or time-unvarying with respect to the $k$th month ($k=1,\ldots,K$): $X_{1,ijk}$ was a continuous variable independently following $\text{Normal}(2+M_{1,ijk},0.5)$; $X_{2,ijk}$ was a binary variable independently following $\text{Binomial}(1,\{1+\exp(-M_{1,ijk}+ M_{2,ijk})\}^{-1})$; $X_{3,ijk}$ following $\text{Normal}(M_{1,ijk}, 0.5)$ was a baseline variable that remained the same over months, e.g. height or gender; similarly, $X_{4,ijk}$ and $X_{5,ijk}$ were time-unvarying variables following $\text{Normal}(1, 1)$ and  $\text{Binomial}(1,0.5)$, respectively. Together with all the generated variables, the underlying model of generating the main outcome was 
$$Y_{ijk}=X_{1,ijk}+0.5X_{2,ijk} -0.5X_{3,ijk} + 0.5X_{4,ijk} + X_{5,ijk} + M_{1,ijk}+\epsilon_{ijk},$$ where the residual vector $\bfepsilon_{ij}$ \red{in Table \ref{biashigh}} followed multivariate normal distribution with mean zeros, variance ones, and a correlation matrix following an unstructured correlation structure with the correlation coefficient between two closer visits being higher (the detailed correlation matrix is presented in Section 2 of the Supplementary Materials). \red{The residual vector $\bfepsilon_{ij}$ in Table \ref{biashighwithgamma} followed a mixture of normal and gamma distributions with a zero mean. The detailed specification and correlation matrix are presented in Section 2 of the Supplementary Materials. The latter setup accounts for the situation of heavily skewed outcomes in healthcare studies \cite{lee2019evaluation}.}

After that, we considered the following missing \revisionone{data} mechanism that was commonly adopted in literature \cite{robins1995analysis,chen2019empirical}: the truncation by death model, i.e., for $k=2,\ldots,K$, $\lambda_{ijk}=Pr(R_{ijk}=1|R_{ij(k-1)}=1,\tilde{\mathbf{H}}_{ijk},\bfalpha)=\{1+\exp(-\alpha_0-X_{1,ijk}+Y_{ij(k-1)})\}^{-1}$, was controlled by the subject-level covariate $X_{1,ijk}$ and the lagged outcome $Y_{ij(k-1)}$ (observed outcome in the most previous month), which satisfied the mechanism of MAR \cite{liang1986longitudinal,robins1995analysis,casals2014methodological,shardell2018joint, chen2019empirical,chen2021multiple}. The parameter $\alpha_0$ was tuned to achieve different missing rates ($\alpha_0=0.5$ for $30\%$ missing rate and $\alpha_0=-0.5$ for $15\%$ missing rate). Finally, the observations after the $(k-1)$th month would be missing if $R_{ijk}=0$. Noted that the first observation was always observed, i.e., $R_{ij1}=1$ for all $i$ and $j$.
 
\subsection{Result assessment}\label{Result assessment} 
Results were based on the weight matrix specified by $\hat{\bfW}_1=Diag(\bfR/\hat{\bfpi})$. The results based on $\hat{\bfW}_2=Diag(\bfR/\hat{\bfpi}^\ast)$ have been presented in Section 2 in the Supplementary Material. We performed extensive simulation studies to evaluate the finite-sample performance of our proposed method. Table \ref{biashigh} , \red{Table \ref{biashighwithgamma}, and Table \ref{sdhigh}} were the results based on 30\% of total participants who were truncated by death, while we considered four scenarios for this set-up: 1) $100$ hospitals with an average $60$ patients per hospital; 2) 100 hospitals with an average $25$ patients per hospital; 3) $25$ hospitals with an average 60 patients per hospital; 4) $25$ hospitals with an average $25$ patients per hospital, \red{with Normal distributed (Table \ref{biashigh}) and non-Normal distributed residuals (Table \ref{biashighwithgamma}).} The case with \red{a larger number of} hospitals would be considered and discussed \red{in Table \ref{kmeans}}. For each scenario, $500$ Monte Carlo replicates were generated with summary statistics of the parameter estimates including absolute bias and Monte Carlo standard derivation (MCSD). In addition,  the asymptotic standard error (ASE) and $95\%$ coverage probability (CP) were evaluated for our method (Table \ref{sdhigh}). 

All values listed in Table \ref{biashigh}, \red{Table \ref{biashighwithgamma}, and Table \ref{sdhigh}} were multiplied by $100$ to \red{avoid digit keeping}. We compared the proposed estimator with four existing estimators MM, GEE, WGEE, CWGEE described in Section \ref{Existing method}. Notice that the MM estimator was a mixed-effect model based estimator with two random effects that was observed not converged in $10\%$ of $500$ Monte Carlo runs, \revisionone{highlighting its numerical limitation in practice}. Thus, we only presented results of MM by excluding NA estimates. From \red{both Table \ref{biashigh} and Table \ref{biashighwithgamma}}, we observed that the bias of existing methods could be hundreds (or thousands) of times larger than the bias of our method. \revisionone{Regardless of residual distributions and the number of hospitals and subjects per hospital}, our estimator only showed negligible bias across different scenarios. The failure of the MM estimator could be attributed to its parametric specification, which was violated in our setting. We would expect similar behaviors for more complex random effect-based models, including joint modeling \cite{shen2021joint}. \revisionone{We also note that existing methods introduce bias in all parameter estimations, not just limited to the intercept estimate. }

Table \ref{sdhigh} summarizes and compares the MCSD and ASE, as well as the coverage rate based on 95\% confidence intervals for our estimator (Omni) \red{with the proposed K-means procedure (refer to Section \ref{Asymptotic property}). We also evaluated the MCSD, ASE and CP for Omni estimator without K-means in Supplementary Material (Table S.4) and observed that the CPs were slightly better after applying the proposed K-means procedure. We also observed that} the derived ASE became closer to MCSD as the number of subjects in one hospital increased. The coverage probabilities were \red{observed to be} satisfactory (around the $95\%$ nominal level) in all scenarios. 

More simulation results are presented in Section 2 of the Supplementary Material, where we additionally considered a \revisionone{truncation rate of around $15\%$} (Table S.1 and Table S.2) and \revisionone{an alternative weight matrix specified by $\hat{\bfW}_2=Diag(\bfR/\hat{\bfpi}^\ast)$ (Table S.3)}. Similar patterns were detected. We found that the results based on $\hat{\bfW}_1$ were slightly better than the results based on $\hat{\bfW}_2$, in terms of coverage probability (Table S.3). Thus, we advocate the use of $\hat{\bfW}_1$ in our application.

In addition to the above scenarios, we also considered the case where the hospital number was much larger ($250$ or $500$ hospitals), but each hospital had a very small sample size (below 10 patients). Table \ref{kmeans} summarizes the results with and without utilizing an unsupervised clustering machine learning algorithm, \revisionone{such as K-means with auto-detected cluster size (refer to Section \ref{Asymptotic property} for more details)}. We observed that all results show little estimation bias, which matches our expectation in Section \ref{Asymptotic property}. However, ASE tended to underestimate MCSD for the estimate without applying clustering, resulting in a coverage probability much lower than its nominal level. By contrast, the estimate with a clustering algorithm tended to better estimate MCSD and lead to a coverage probability closer to $95\%$. In reality, many hospitals may have the same unknown features owing to close geographic locations or shared policies and culture across states, so it is quite reasonable to assume that there are limited latent categories among hospitals. \red{Moreover, not limited to Silhouette score,} we can advocate other useful tools in the literature to further guide the selection of the cluster number\cite{pinheiro2014model}.

\begin{table}
\caption{To detect patient-level factors associated with recovery of ADRD older adult after hip fracture. The values below are effect sizes with $*$ indicating significance (P-value$<0.05$). MM: mixed-effect model; Omni: our proposed method; LOS\_DAY: length of stay days in hospital; ELIXHAUSER\_SCORE: multimorbidity counts within half a year before hip fracture; PRE\_DAH: DAH in the month right before hip fracture.}
\label{realmain}
\begin{tabular}{llccccc}
\toprule
            &                   & MM   & GEE    & WGEE   & CWGEE  & Omni   \\ \midrule
BED\_SIZE$<230$ & SEX               & -0.03  & -0.39  & -0.49  & -0.97  & 0.04   \\
            & LOS\_DAY          & -0.42* & -0.43* & -0.47* & -0.53* & -0.44* \\
            & ELIXHAUSER\_SCORE & -0.15* & -0.03  & 0.01   & -0.14  & -0.05  \\
            & AGE$(76-85)$              & -1.85* & -1.4*  & -1.33* & -1.27  & -1.47* \\
            & AGE$(85+)$              & -3.09* & -2.76* & -3.42* & -3.28* & -3.15* \\
            & MONTH             & 10.64* & 10.5*  & 10.26* & 10.82* & 10.25* \\
            & MONTH\_SQUARE     & -1.1*  & -1.06* & -1.05  & -1.11* & -1.05* \\
            & PRE\_DAH          & 0.18*  & 0.25*  & 0.32*  & 0.29*  & 0.29*  \\ \midrule
$231\leq$BED\_SIZE$<424$ & SEX               & 0.45   & 0.58*  & 1.08*  & 0.91   & 0.87*  \\
            & LOS\_DAY          & -0.39* & -0.4*  & -0.43* & -0.44* & -0.44* \\
            & ELIXHAUSER\_SCORE & -0.2*  & -0.12  & -0.08  & -0.14  & -0.2*  \\
            & AGE$(76-85)$              & -1.66* & -1.30* & -1.55* & -0.04  & -1.23* \\
            & AGE$(85+)$              & -2.81* & -2.3*  & -2.71* & -1.43* & -2.59* \\
            & MONTH             & 11.55* & 11.39* & 11.12* & 11.37* & 11.05* \\
            & MONTH\_SQUARE     & -1.2*  & -1.15* & -1.13* & -1.16* & -1.12* \\
            & PRE\_DAH          & 0.19*  & 0.24*  & 0.31*  & 0.31*  & 0.27*  \\ \midrule
BED\_SIZE$\geq 424$ & SEX               & -0.18  & -0.31  & -0.09  & 1.04   & -0.23  \\
            & LOS\_DAY          & -0.3*  & -0.3*  & -0.16* & -0.35* & -0.18* \\
            & ELIXHAUSER\_SCORE & -0.31* & -0.22* & -0.32* & -0.46* & -0.38* \\
            & AGE$(76-85)$              & -1.00* & -0.81  & -0.89  & -1.24  & -0.8   \\
            & AGE$(85+)$              & -2.01* & -1.69* & -2*    & -2.21* & -1.93* \\
            & MONTH             & 11.53* & 11.38* & 11.15* & 11.88* & 11.11* \\
            & MONTH\_SQUARE     & -1.19* & -1.15* & -1.13* & -1.23* & -1.12* \\
            & PRE\_DAH          & 0.18*  & 0.23*  & 0.28*  & 0.23*  & 0.25* \\ \bottomrule
\end{tabular}
\end{table}

\section{Real data application}\label{A real data application}
Operationalization of the post-discharge outcome (DAH) was motivated by the health services research literature \cite{groff2016days,burke2020healthy}. Using Medicare claims for each month after hospital discharge, we calculated the number of days spent in hospitals, nursing homes, or rehabilitation facilities by subtracting them from $30$. We also subtracted days with emergency department visits or observation days. This DAH metric is a novel population-based outcome measure developed in conjunction with the Medicare Payment Advisory Commission \cite{burke2020healthy}. To better design and organize the care in ways that improve outcomes of older adults with ADRD, our goal in this study was to examine how DAH varied by ADRD beneficiary demographic characteristics as well as healthcare market among a random sample of Medicare fee-for-service beneficiaries (5\% random sample for 2010 to 2016, 20\% random sample for 2017) who were aged 65 years or older and had experienced a traumatic hip fracture. Hip fractures were identified using ICD-9 and ICD-10 codes consistent with femoral neck or intertrochanteric fractures, excluding fractures of the pelvic rim or acetabulum \cite{lee2019evaluation}. From the selection criteria in our study, a few participants were excluded if they were not community dwelling at baseline (e.g., were admitted from nursing homes), died during the index hip fracture hospital admission, or were not continuously eligible for Medicare fee-for-service for 6 months prior and 12 months following the fracture. We also excluded a very small number of patients who were discharged against medical advice. As a result, we had 16,562 ADRD patients from 2,268 hospitals after hospital discharge, among whom 4,126 (24.9\%) died during the half-year follow-up post-discharge. Furthermore, we took into account \revisionone{clinically meaningful} patient-specific factors that were identified by the \revisionone{epidemiologist and }gerontologist as the most relevant to the post-acute recovery of older adults, such as length of stay days in hospital (LOS\_DAY), multimorbidity counts within half a year before hip fracture (ELIXHAUSER\_SCORE), sex (0: male; 1: female), age group (0: 65-75; 1: 76-85; 2: 85+), month (linear trend), quadratic trend (MONTH\_SQUARE), and DAH in the month right before hip fracture (PRE\_DAH) as our primary interest in the analysis. 

\begin{figure}
    \centering
    \includegraphics[width=6in]{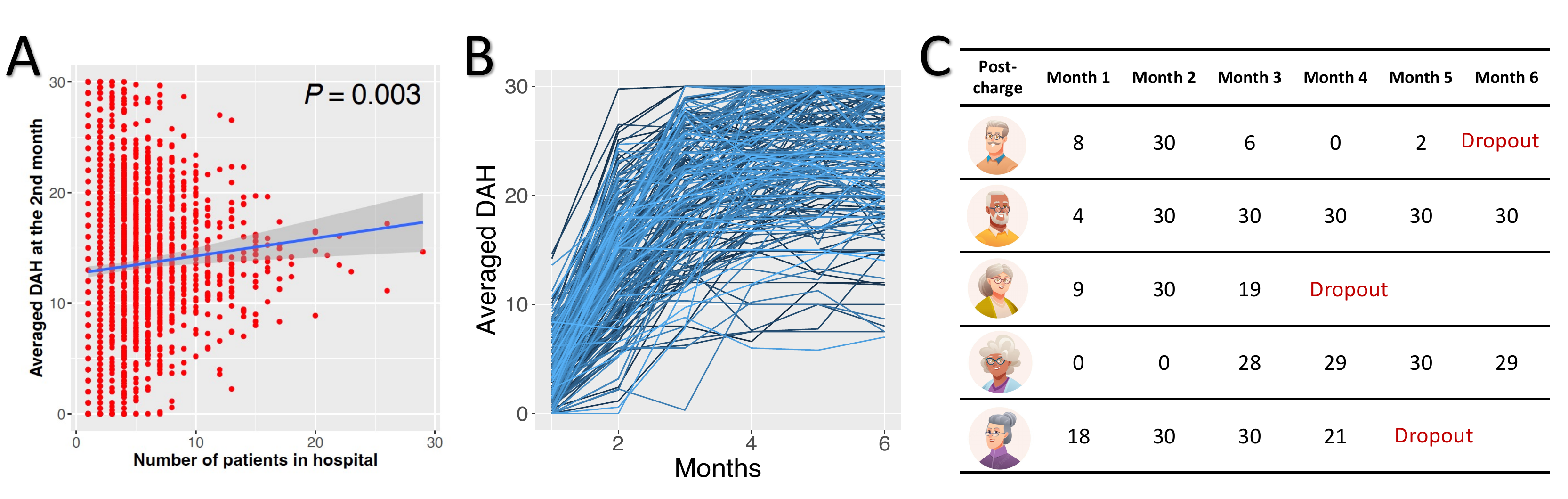}
    \caption{A. Informative Cluster Size with P-value calculated by a simple linear model. B. Hospital-level variability: each trajectory represents averaged DAH at each month. C. Data visualization.}
    \label{real data pic}
\end{figure}

\revisionone{Before conducting analysis, we evaluated the existence of three challenges in Medicare administrative healthcare claims data, namely patient truncation by death (Figure \ref{real data pic}C), potential ICS for each hospital (Figure \ref{real data pic}A), and hospital-level unmeasured confounding (Figure \ref{real data pic}B)}. The presence of any of these three issues can lead to biased estimation and thus invalid statistical inference. Figure \ref{real data pic}A shows that the number of older adults treated in each hospital is positively associated with DAH at months 2 and 3 after hospital discharge \revisionone{(P-value=$0.003$)}. This observation suggests that hospitals with more patients may lead to better rehabilitation outcomes, possibly through the exploitation of economies of scale. \revisionone{On the other hand,} we statistically modeled truncation by death using DAH from the last observed month as a risk factor, as shown in \red{Table S.5} in the Supplementary Material, which indicates that this variable has a highly significant increasing trend. This finding provides evidence that the death process is not MCAR \cite{bhaskaran2014difference}. The MAR mechanism was plausible, given that the previous month's observed DAH was a good surrogate outcome and a strong indicator of post-discharge mortality \cite{zogg2022beyond}. As a result, we assumed that the truncation by death mechanism was independent of DAH conditional on the last observed DAH and other variables \cite{bhaskaran2014difference}. \revisionone{In addition, we evaluated potential lack of fit of missing data model using deviance and chi-square test. The resulting P-value is close to one, which implies that the missing data model may fit well.} Moreover, the visualization in Figure \ref{real data pic}B suggests a large variability of averaged DAH across hospitals. This finding may be explained by the presence of unmeasured hospital-level confounders that are not sufficiently explained by observed data. \revisionone{Accordingly, these three challenges may indeed be present in this data, underscoring the importance of implementing our method, as showcased in earlier sections.}

We applied our proposed estimator (Omni) and compared it with other existing methods, such as WGEE, CWGEE, and MM. \revisionone{To account for observed heterogeneity of hospital profiles due to scale efficiency and optimal size of the hospital sector \cite{giancotti2017efficiency}, we conducted a stratified analysis based on a hospital-level score called ``bed size"} \cite{giancotti2017efficiency}, which was valued as 0, 1, or 2. These categories represented small hospitals ($<230$ beds, bed size = 0), medium-sized hospitals ($231$ to $424$ beds, bed size = 1), or large hospitals ($>425$ beds, bed size = 2). A larger "bed size" indicated a bigger hospital capacity with richer medical resources and more experts. This variable was defined using publicly available Medicare provider data from the Provider of Services file. Additionally, we used K-means with Silhouette score \cite{rousseeuw1987silhouettes} to identify the best cluster number for $\hat{\bftheta}$. In Figure S.2 in the Supplementary Material, we detected two latent patterns among hospitals with bed size equal to 0 or 1, while five latent patterns were detected among hospitals with bed size equal to 2. Thus, more heterogeneity in the data might exist among large-sized hospitals.

From Table \ref{realmain}, we observed that different methods yielded slightly different conclusions regarding factors related to DAH. The findings common to all estimators were that 1) older age and longer LOS\_DAY in hospital were significantly associated with fewer DAH after hospital discharge, while longer time after hip fracture and more PRE\_DAH before hip fracture were associated with more DAH; 2) comparing hospitals with different bed sizes, we found that larger bed size was associated with more DAH. This result could be explained by the fact that larger hospitals have more medical resources and greater expertise in the clinical management of hip fracture survivors, leading to quicker recovery after hospital discharge. However, only the GEE, WGEE, and Omni methods showed significantly better recovery for females in medium-sized hospitals, and GEE, WGEE, CWGEE, and Omni methods showed no significant difference in effects on DAH when comparing the 65-to-75-year-old group and the 76-to-85-year-old group in medium-sized hospitals. We also found that a larger hospital “bed” size was associated with a more negative effect of comorbidity count on DAH.


\section{Discussion}\label{discussion}
\revisionone{We have developed a useful and valid toolbox by  creatively adapting and modifying existing mathematical and data science tools to handle three sources of statistical complexity in analyzing Medicare claims, while existing commonly-used methods fail to work and result in significant estimation bias. Our numerical evaluations affirm the superiority of our method over existing ones and validate the theoretical derivations}. 

Our data application highlight the importance of hospitals, healthcare providers, and companies paying closer attention to beneficiaries with older age (especially aged $\geq 85$ years), longer LOS\_DAY in the hospital, and more chronic health conditions and \revisionone{less days spent at home} before hip fracture. This group of older adults is more likely to have a worse recovery after hospital discharge. Our findings also suggest that society and the healthcare system should focus on smaller-sized hospitals by allocating medical resources to better treat patients. Additionally, we observed potential association between bed size and the effect size of the comorbidity count. This could be explained by the fact that larger hospitals often treat sicker patients with severe health conditions. This finding aligns with our speculation on hospital-level heterogeneity and motivates future work to identify hospital-level factors driving the differences. The above findings provide informative insights and have the potential to help society and the healthcare system reorganize, allocate resource, and improve the care of older adults with Alzheimer's disease and related dementias (ADRD). Compared to those without ADRD, older adults with ADRD are at a much higher risk of experiencing hip fracture, yet optimal clinical management remains unclear. Our application highlights the importance of taking more urgent action to improve the recovery of hip fracture survivors living with ADRD.

 \revisionone{In addition to the content presented in the manuscript, in Section 3 of the Supplementary Material, we have introduced an extension designed for scenarios involving high-dimensional covariates within the main model. Furthermore, in Section 4 of the Supplementary Material, we have delved into an extension for managing time-varying unmeasured hospital-level confounders/effects.}  Comprehensive theoretical and numerical assessments are required to gauge their effectiveness and practical utility.


Despite what we have achieved, there are several limitations in the current framework, which merit more efforts in future studies. First, the significant factors of beneficiaries detected by our model may not be causal. To further make a solid argument about causality, a causal inference model should be adopted \cite{robins2000marginal,chen2023multiple}; second, the current framework cannot make any inference on hospital-level factors as a result of PLS, even when these variables are observed. However, we can still make inference in our model on interactions of patient-level factors and observed hospital-level factors; third, the assumption of correctly specified truncation by death model may not hold in some applications. More advanced tool in theory such as doubly robust estimation can be considered \cite{bang2005doubly}; last but not the least, additional data are needed to explore spatial heterogeneity and \revisionone{biomarkers effects (e.g., neuroimaging biomarkers describing the illness of ADRD )} among beneficiaries in different communities, which is highly likely to have strong effect on recovery of ADRD beneficiaries with hip fracture. All deserve substantial effort and merit future work.

\nocite{*}
\bibliography{wileyNJD-AMA}%

\section*{Acknowledgements}
This work was supported by Maryland Claude D. Pepper Center (P30AG028747), National Institute on Aging (K76AG074926), and National Center for Advancing Translational Sciences (NCATS) and Clinical Translational Science Award (CTSA) (1UL1TR00309)

\subsection*{Author contributions}
All authors have made important contributions and have approved this work. 

\subsection*{Financial disclosure}

None reported.

\subsection*{Conflict of interest}

The authors declare no potential conflict of interests.

\subsection*{Data Availability Statement}
The data that support the findings of this study are available upon reasonable request which should be directed to the Dr. Chixiang Chen and Dr. Jason Falvey at University of Maryland School of Medicine. 

\end{document}